\documentclass[
reprint,
superscriptaddress,
amsmath,amssymb,
aps,
prl,
]{revtex4-1}

\usepackage{graphicx}
\usepackage{dcolumn}
\usepackage{upgreek}
\usepackage{mathrsfs}
\usepackage{setspace}
\usepackage{bm}
\usepackage[breaklinks=true,colorlinks=true,linkcolor=blue,urlcolor=blue,citecolor=blue]{hyperref}
\allowdisplaybreaks[4]
\begin{document}

\preprint{APS/123-QED}

\title{Quantum Transport of Rydberg Excitons with Synthetic Spin-Exchange Interactions}
\author{Fan Yang}
\affiliation{State Key Laboratory of Low Dimensional Quantum Physics, Department of Physics, Tsinghua University, Beijing 100084, China}
\author{Shuo Yang}
\email{shuoyang@mail.tsinghua.edu.cn}
\affiliation{State Key Laboratory of Low Dimensional Quantum Physics, Department of Physics, Tsinghua University, Beijing 100084, China}
\author{Li You}
\email{lyou@mail.tsinghua.edu.cn}
\affiliation{State Key Laboratory of Low Dimensional Quantum Physics, Department of Physics, Tsinghua University, Beijing 100084, China}
\affiliation{Beijing Academy of Quantum Information Sciences, Beijing 100193, China}


\begin{abstract}
We present a scheme for engineering quantum transport dynamics of spin excitations in a chain of laser-dressed Rydberg atoms, mediated by synthetic spin-exchange arising from diagonal van der Waals interaction. The dynamic tunability and long-range interaction feature of our scheme allows for the exploration of transport physics unattainable in conventional spin systems. As two concrete examples, we first demonstrate a topological exciton pumping protocol that facilitates quantized entanglement transfer, and secondly we discuss a highly nonlocal correlated transport phenomenon which persists even in the presence of dephasing. Unlike previous schemes, our proposal requires neither resonant dipole-dipole interaction nor off-diagonal van der Waals interaction. It can be readily implemented in existing experimental systems.
\end{abstract}

\maketitle

Developing controlled large-scale quantum systems constitutes a central goal of quantum simulation and quantum computation \cite{cirac2012goals,georgescu2014quantum}. Among the variety of physical realizations, neutral atoms present several unique advantages \cite{weiss2017quantum}, their inherent qubit identity, long coherence time, flexible state maneuverability, as well as tunable qubit-qubit interactions, for instance mediated by Rydberg states \cite{saffman2010quantum}. The continued progresses in Rydberg-atom studies offer great potential for probing many-body dynamics \cite{labuhn2016tunable,bernien2017probing,marcuzzi2017facilitation,kim2018detailed,guardado2018probing}. With improved operation fidelity \cite{levine2018high} and increased system size \cite{barredo2018synthetic}, quantum simulation on the Rydberg atom based platform \cite{kim2018detailed,takei2016direct} is becoming increasingly attractive.

The transport of particle or spin via quantum state-changing interactions is essential for understanding energy or information flow. Emulating such problems on a quantum simulator constitutes a focused thrust within the broad quantum physics community \cite{vzutic2004spintronics,najafov2010observation,collini2013spectroscopic,jurcevic2014quasiparticle,akselrod2014visualization}. Earlier efforts based on Rydberg-atom systems have provided first insights \cite{barredo2015coherent,schonleber2015quantum,orioli2018relaxation,gunter2013observing,schempp2015correlated,letscher2018mobile,wuster2011excitation}, where transport of spin excitation is facilitated typically by resonant dipole-dipole interaction (DDI) or by off-diagonal van der Waals (vdW) flip-flop interaction between Rydberg states. They include direct spin-exchange between different Rydberg states \cite{barredo2015coherent,schonleber2015quantum,orioli2018relaxation}, second-order exchange between ground state and Rydberg state \cite{gunter2013observing,schempp2015correlated,letscher2018mobile}, and a fourth-order process inside ground internal state manifolds \cite{wuster2011excitation,glaetzle2015designing,van2015quantum}.

In this Letter, we propose a simpler yet as effective method for engineering exciton transport dynamics in a Rydberg-atom system. The use of resonant DDI or flip-flop vdW interaction is avoided. Instead, our main idea relies on a perturbative spin-exchange process by off-resonantly dressing the ground state to a Rydberg state. Capitalizing on the diagonal vdW interaction-induced Rydberg level shift, perturbations from different pathways collectively contribute to a net exchange interaction between the ground and the Rydberg states. When exciton-exciton interaction as well as dephasing are included, our model system is shown to be capable of simulating various transport dynamics unattainable in conventional spin systems. In the first example, we establish an interesting topological pumping protocol, whereby the exciton experiences a quantized center-of-mass motion. In the second example, we show that the long-range interaction between excitons permits the formation of high-order magnon bound state, which exhibits nonlocal correlations even when ballistic transport turns into classical diffusion due to dephasing.

\begin{figure}
\centering
\includegraphics[width=\linewidth]{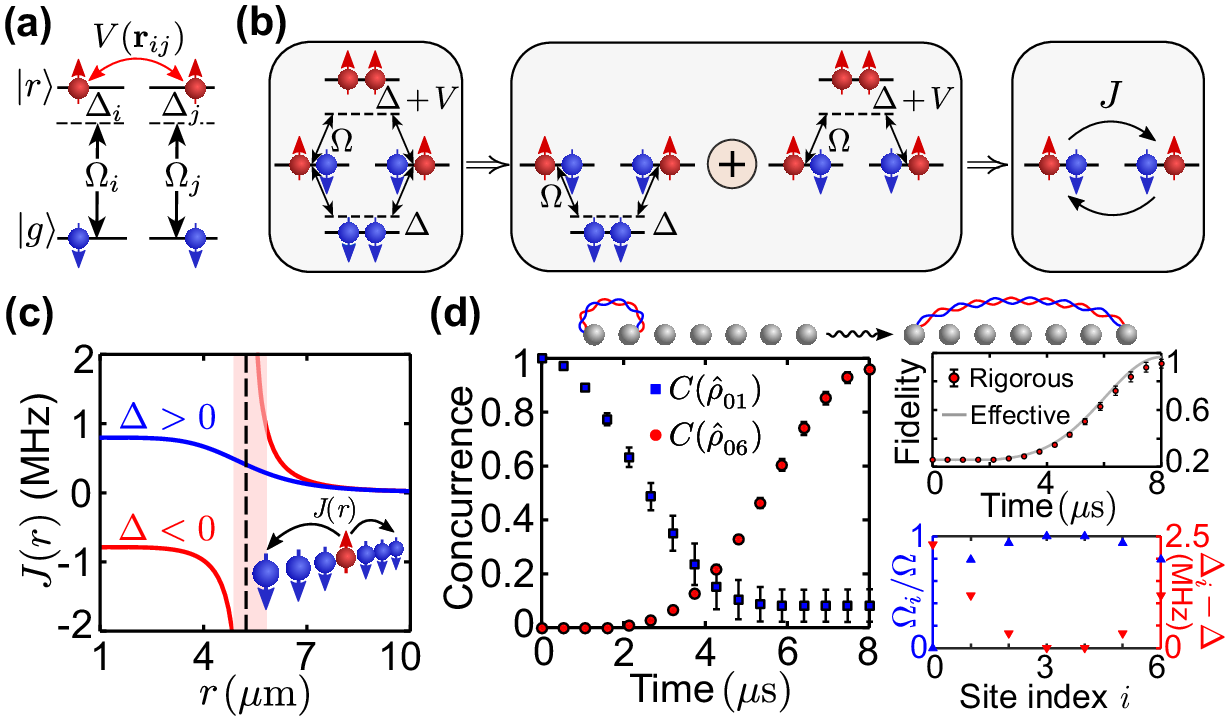}
\caption{(a) Level structure for the proposed atomic system. We consider $^{87}$Rb atom with $|g\rangle=|5S_{1/2},F=2,m_F=-2\rangle$, $|r\rangle=|70S,J=1/2,m_J=-1/2\rangle$ \cite{levine2018high}. (b) Illustration of the mechanism for generating synthetic spin-exchange interaction. (c) Synthetic exchange strength $J(r)$ versus distance $r$ for $\Omega/2\pi=5$~MHz and $\Delta/2\pi=\pm50$~MHz. The dashed lines denote the facilitation condition $\Delta+V(r)=0$. (d) The left panel shows the concurrence for the first two nodes [$C(\hat{\rho}_{01})$] and the two end nodes [$C(\hat{\rho}_{06})$]. The right panel shows the fidelity to the target state $(|r_0g_1\cdots g_6\rangle-i|g_0g_1\cdots r_6\rangle)/\sqrt{2}$ (upper figure, the rigorous and effective results are obtained with $\hat{H}$ and $\hat{H}_\mathrm{eff}$, respectively) and values of dressing parameters (lower figure) with $V(d)=3\Delta$ ($\Delta>0$) and $d=4.4~\mu\mathrm{m}$. The numerical data represent averages over 500 calculations, assuming a Gaussian distribution of atomic position along the chain direction with $0.1~\mu\mathrm{m}$ standard deviation. The error bars mark one standard deviation intervals.}
\label{fig:fig1}
\end{figure}

{\it Model}.---The system we study is an array of individually trapped cold atoms, dressed by laser fields that couple the ground state $|g\rangle$ to a Rydberg state $|r\rangle$ \cite{marcuzzi2017facilitation,levine2018high}. It is modeled by the Hamiltonian
\begin{equation}
\label{eq:eq1}
\hat{H}=\sum_{i}\frac{\Omega_i}{2}\hat{\sigma}_x^i+\sum_i\Delta_i\hat{\sigma}_{rr}^i+\sum_{i<j}V(\mathbf{r}_{ij})\hat{\sigma}_{rr}^i\hat{\sigma}_{rr}^j,
\end{equation}
where $\Omega_i$ and $\Delta_i$ are Rabi frequencies and detunings of the dressing field [Fig.~\ref{fig:fig1}(a)], $\hat{\sigma}_x^i=|r_i\rangle\langle g_i|+|g_i\rangle\langle r_i|$ and $\hat{\sigma}_{\alpha\alpha}^i=|\alpha_i\rangle\langle\alpha_i|$ ($\alpha=g,r$) are spin-flip and projection operators for the $i$-th atom (located at $\mathbf{r}_i$), and $V(\mathbf{r}_{ij})=C_6/|\mathbf{r}_i-\mathbf{r}_j|^6$ is the diagonal vdW interaction between atoms in the Rydberg state \cite{bernien2017probing} (we take $C_6>0$).

First, we consider the dynamics of a single Rydberg exciton. In the limit of large detuning with $\Omega_i\ll|\Delta_i|$, $|\Delta_i+V(\mathbf{r}_{ij})|$; and $|\Delta_i-\Delta_j|\ll|\Delta_{i/j}|$, the singly excited state set $\{|\Psi_i\rangle=|g_1g_2\cdots r_i\cdots g_N\rangle,i=1,2,\cdots N\}$ forms a quasi-degenerate subspace $\Pi_1$. As a result, the perturbative coupling with the rest of the Hilbert space can induce strong state mixing inside $\Pi_1$, giving rise to coherent position exchanges of the exciton. To clarify the basic physics, we take the example of $N=2$. As shown in Fig.~\ref{fig:fig1}(b), the degenerate states $|r_1g_2\rangle$ and $|g_1r_2\rangle$ are off-resonantly coupled to $|g_1g_2\rangle$ and $|r_1r_2\rangle$, which can be approximately treated as two Raman pathways. For the non-interacting case ($V=0$), the contributions of these two paths cancel out. In the presence of the vdW interaction, the level shift $V$ for $|rr\rangle$ causes the two pathways collectively to yield a nonvanishing spin exchange interaction $J=\Omega^2/4\Delta-\Omega^2/4(\Delta+V)$. For a many-body system, applying second-order van Vleck perturbation theory \cite{shavitt1980quasidegenerate} to the original model [Eq.~(\ref{eq:eq1})] and dropping the constant $-\sum_{j}\Omega_j^2/4\Delta_j$, we arrive at an effective Hamiltonian \cite{supply}
\begin{equation}
\label{eq:eq2}
\hat{H}_\mathrm{eff} =\sum_{i}\left(\Delta_i+\frac{\Omega_i^2}{2\Delta_i}\right)\hat{\sigma}_{rr}^i + \sum_{i\neq j} I_{ij}\hat{\sigma}_{rr}^i\hat{\sigma}_{gg}^j + J_{ij}\hat{\sigma}_{+}^i\hat{\sigma}_{-}^j,
\end{equation}
where $\hat{\sigma}_{+}^i=|r_i\rangle\langle g_i|$ and $\hat{\sigma}_{-}^i=|g_i\rangle\langle r_i|$ are spin raising and lowering operators for the $i$-th atom. The Ising-type interaction $I_{ij}$ and the spin-exchange interaction $J_{ij}$ respectively take the following forms
\begin{equation}
I_{ij} = \frac{\Omega_j^2V(\mathbf{r}_{ij})}{4\Delta_j[\Delta_j+V(\mathbf{r}_{ij})]},
J_{ij} = \sum_{\beta=i,j}\frac{\Omega_i\Omega_jV(\mathbf{r}_{ij})}{8\Delta_\beta[\Delta_\beta+V(\mathbf{r}_{ij})]}.\nonumber
\end{equation}
In contrast to earlier dressing schemes \cite{glaetzle2015designing,van2015quantum}, the spin-exchange interaction we find constitutes a pure synthetic interaction as the initial Hamiltonian Eq.~(\ref{eq:eq1}) contains only diagonal vdW interactions. It exhibits different $r$-dependence compared with previous schemes [Fig.~\ref{fig:fig1}(c)] and is highly tunable in terms of $\Omega_i$ and $\Delta_i$. This effective model is not restricted to any particular type of lattice, and this work considers the simplest one-dimensional periodic chain with a spacing $d$.


The exciton transport is conveniently described by mapping spins to hard-core bosons with $\hat{\sigma}_+^i=\hat{a}_i^\dagger$ and $\hat{\sigma}_-^i=\hat{a}_i$, where $\hat{a}^\dagger_i$ ($\hat{a}_i$) creates (annihilates) a Rydberg exciton at site $i$. The effective Hamiltonian for a single exciton can then be expressed in a tight-binding form with $\hat{H}_\mathrm{eff}=\sum_i\mu_i\hat{a}^\dagger_i\hat{a}_i+\sum_{i<j}J_{ij}\left(\hat{a}^\dagger_i\hat{a}_{j}+\hat{a}^\dagger_j\hat{a}_{i}\right)$, where $\mu_i=\Delta_i+\Omega_i^2/2\Delta_i+\sum_{j\neq i}I_{ij}$ is the on-site potential. To benchmark the effective model, we focus on an entanglement distribution protocol, in which the entangled state $(|r_0g_1\rangle+|g_0r_1\rangle)/\sqrt{2}$ is transferred to $(|r_0g_N\rangle+e^{i\phi}|g_0r_N\rangle)/\sqrt{2}$ over a chain of $N+1$ atoms. For systems dominated by nearest-neighbor (NN) hopping, perfect entanglement transfer can be achieved when the conditions $\mu_i=\mu$, $J_{i,i+1}=J\sqrt{i(N-i)}$ \cite{christandl2004perfect} are satisfied. The initial entangled state can be set up via Rydberg blockade, and the perfect transfer condition can be met by tuning local parameters $\Omega_i$ and $\Delta_i$. As verified by numerical results ($N=6$) shown in Fig.~\ref{fig:fig1}(d), entanglement between the two end nodes gradually establishes and approaches the maximal value eventually, calibrated by their concurrence \cite{wootters1998entanglement} and state fidelity. The potential existence of disorder in atomic positions is also taken into account in the calculation \cite{marcuzzi2017facilitation}. As long as the disorder-induced interaction fluctuation $\delta V_{ij}$ is much smaller than $V_{ij}$ itself ($\delta V_{ij}\ll V_{ij}$), the transport efficiency remains high \cite{supply}. We note that all numerical results presented in this work are based on solving the exact model Eq.~(\ref{eq:eq1}).

In addition to simulating coherent dynamics, the system we consider also provides a platform for probing the crossover between coherent and incoherent transport. In the presence of dephasing \cite{lesanovsky2013kinetic,lienhard2018observing}, the evolution of the density matrix $\hat{\rho}$ is governed by the master equation $\partial_t{\hat{\rho}} =-i[\hat{H},\hat{\rho}]+\sum_{i}\mathcal{L}[\sqrt{\gamma}\hat{\sigma}_{rr}^{i}]\hat{\rho}$, where the Lindblad operator $\mathcal{L}$ gives $\mathcal{L}[\hat{\sigma}]\hat{\rho}=\hat{\sigma}\hat{\rho}\hat{\sigma}^\dagger-\frac{1}{2}(\hat{\sigma}^\dagger\hat{\sigma}\hat{\rho}+\hat{\rho}\hat{\sigma}^\dagger\hat{\sigma})$. For weak dephasing ($\gamma\ll|\Delta_i|$), the dynamics remain confined within the subspace $\Pi_1$ for times smaller than $t_c=\min\left\{\Delta_i^2/\gamma\Omega_i^2\right\}$, while for $t>t_c$ incoherent spin flips and exciton growth takes over \cite{lesanovsky2013kinetic}. In the transport regime ($t<t_c$), the dynamics can be effectively described by $\partial_t{\hat{\rho}} =-i[\hat{H}_\mathrm{eff},\hat{\rho}]+\sum_{i}\mathcal{L}[\sqrt{\gamma}\hat{a}_i^\dagger\hat{a}_i]\hat{\rho}$, equivalent to the Haken-Reineker-Strobl (HRS) model with coherent hopping and on-site dephasing \cite{haken1972coupled,haken1973exactly,reineker1973mean}. The exciton motion remains coherent for $t<1/\gamma$, and exhibits incoherent features as it enters the diffusion region $t>1/\gamma$ \cite{supply}.

Next we discuss the case involving multi-excitons. With $n$ Rydberg excitons, the quasi-degenerate perturbation analysis can no longer be simply applied to the subspace $\Pi_n$ spanned by the states $\{|\Psi_{i_1,\cdots,i_n }\rangle=|g_1\cdots r_{i_1}\cdots r_{i_n}\cdots g_N\rangle,i_1<\cdots<i_n\}$, since the large vdW interaction between excitons removes some of the degeneracy. For example, in the $n=2$ case, the doubly excited subspace $\Pi_2$ can be decomposed into $\Pi_2=\Pi^\prime_2\bigcup\Pi^{\prime\prime}_2$, with $\Pi^\prime_2$ spanned by the dimer states $\{|\Psi_{i,i+1}\rangle=|g_1\cdots r_{i}r_{i+1}\cdots g_N\rangle,i=1,\cdots,N\}$ and $\Pi^{\prime\prime}_2$ the complementary set of $\Pi^\prime_2$. If $V(\mathbf{r}_{i,i+1})$ is of the same order as $|\Delta_i|$, $\Pi^{\prime}_2$ and $\Pi^{\prime\prime}_2$ forms two decoupled quasi-degenerate subspaces. Inside $\Pi^{\prime}_2$, the dimer states are coupled to each other through next-nearest-neighbor (NNN) hoppings $\sum_iJ^{(2)}_i\hat{\sigma}^i_+\hat{\sigma}^{i+1}_{rr}\hat{\sigma}^{i+2}_-+\mathrm{H.c.}$, with an effective three-body exchange interaction
\begin{equation}
J_{i}^{(2)} \approx \frac{\Omega_i\Omega_{i+2}V(\mathbf{r}_{i,i+2})}{4[\Delta_i+V(\mathbf{r}_{i,i+1})][\Delta_i+V(\mathbf{r}_{i,i+1})+V(\mathbf{r}_{i,i+2})]}.\nonumber
\end{equation}
For exciton dynamics inside $\Pi^{\prime\prime}_2$, the perturbation analysis yields an effective Hamiltonian $\hat{H}_\mathrm{eff}^\prime=\hat{H}_\mathrm{eff}+\hat{H}_\mathrm{int}$ \cite{supply}, where $\hat{H}_\mathrm{eff}$ is the single-exciton effective Hamiltonian, while $\hat{H}_\mathrm{int}=\sum_{i<j}U_{ij}\hat{a}^\dagger_i\hat{a}^\dagger_j\hat{a}_j\hat{a}_i$ describes exiciton-exciton interactions, with strength $U_{ij}=V(\mathbf{r}_{ij})-2(I_{ij}+I_{ji})$. For the case of $n$ excitons, the dynamics of the system can still be approximately described by $\hat{H}_\mathrm{eff}^\prime$, as long as the initial separations between excitons are large enough to ensure that their mutual interactions are much smaller than the detuning. If some of the excitons are close to each other initially, they will form tightly bound state (such as the dimer state described above), whose transport property needs further elaborations.

When simulating transport physics, we focus on the case where the total exciton number $\hat{N}_e=\sum_{i}\hat{\sigma}_{rr}^i=\sum_{i}\hat{a}_i^\dagger\hat{a}_i$ is conserved. However, $\hat{N}_e$ is not strictly conserved due to a finite spin-flip probability $|\Omega_i/\Delta_i|^2$ for each ($i$-th) atom. This could significantly influence the quality of our simulations especially as the number of atoms increases. However, for observables whose expectation values only depend on the diagonal elements of $\hat{\rho}$ (e.g., density correlations), the simulation results can be refined via post selection based on projective measurements. If the dynamics of $n$ excitons are of interest, the expectation values of the observables are calculated by the refined density matrix $\hat{\rho}_p=p^{-1}\sum_kp_k\hat{P}_k$, where $\hat{P}_k=|\phi_k\rangle\langle\phi_k|$ denotes the projection operator of state $|\phi_k\rangle\in\Pi_n$, and $p_k=\mathrm{Tr}(\hat{\rho}\hat{P}_k)$ is the probability of the measurement. The post-selection probability $p=\sum_k p_k$ scales as $1-N\Omega^2/2\Delta^2$, which is acceptable for a reasonable sized system ($p\approx0.75$ for $\Delta/\Omega=10$ and $N=50$).

\begin{figure}
\centering
\includegraphics[width=\linewidth]{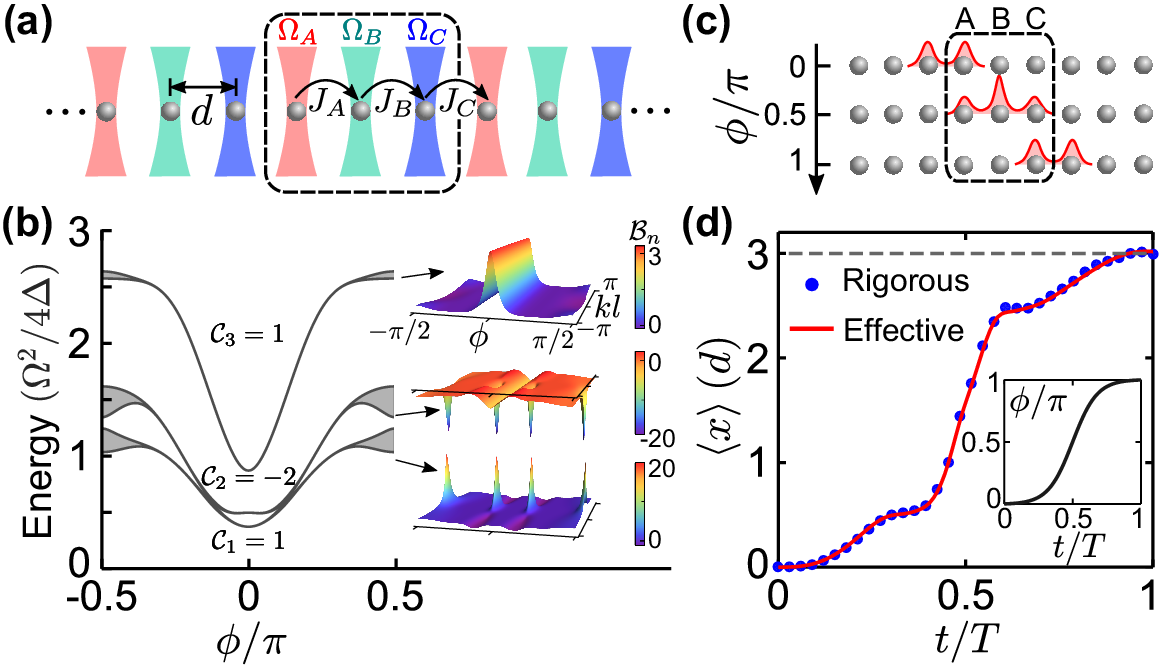}
\caption{(a) Illustration of the dressing scheme for topological exciton pumping. (b) Energy band and Berry curvature of the effective model. (c) Illustration of the pumping sequence. (d) Mean displacement $\langle x\rangle$ of the exciton at different time $t$. The blue dots are calculated by the refined density matrix using the exact model Eq.~(\ref{eq:eq1}), and the red lines are obtained with the effective Hamiltonian Eq.~(\ref{eq:eq3}). The inset shows the modulation detail $\phi/\pi=\frac{1}{2}+\frac{\tanh[5.6(t/T-1/2)]}{2\tanh(2.8)}$. The simulations are performed with $\Omega/2\pi=5$~MHz, $\Delta/2\pi=20$~MHz, $V(d)=3\Delta$, $N=12$, and $T=27.7~\mu\mathrm{s}$.}
\label{fig:fig2}
\end{figure}

{\it Example~1}.---To illustrate the dynamical tunability of our scheme, we consider an implementation for topological exicton pumping, for which a time-dependent and site-dependent exchange interaction is required \cite{thouless1983quantization,qian2011quantum,wang2013topological,lohse2016thouless,nakajima2016topological,lu2016geometrical}. A periodic system with broken parity symmetry is assumed \cite{mueller2004artificial}, with three lattice sites (labeled as $A$, $B$, $C$, and separated by $d$) forming a unit cell (with the period $l=3d$), dressed by control fields of three intensities [Fig.~\ref{fig:fig2}(a)] with corresponding Rabi frequencies $\Omega_A,\Omega_B,\Omega_C=\Omega\times\{\sin^2(\phi+\pi/4),\sin^2(\phi),\sin^2(\phi-\pi/4)\}$ and $\phi$ a time-dependent control parameter. Such a dressing scheme can be realized by using three independently controlled acousto-optic deflectors. Retaining the NN interaction, the system can be described by the generalized Rice-Mele Hamiltonian \cite{rice1982elementary}
\begin{align}
\label{eq:eq3}
\hat{H}_\mathrm{eff}=&\sum_{i}\left(J_A\hat{a}^\dagger_i\hat{b}_{i}+J_B\hat{b}^\dagger_i\hat{c}_{i}+J_C\hat{c}^\dagger_i\hat{a}_{i+1}+\mathrm{H.c.}\right)\nonumber\\
&+\sum_i\left(\mu_A\hat{a}^\dagger_i\hat{a}_i+\mu_B\hat{b}^\dagger_i\hat{b}_i+\mu_C\hat{c}^\dagger_i\hat{c}_i\right),
\end{align}
where $\hat{a}_i$, $\hat{b}_i$, and $\hat{c}_i$ are exciton annihilation operators for site $A$, $B$, and $C$ of the $i$-th unit, respectively. According to the Bloch theorem, this system can be described in the quasi-momentum $k$ space with a single-particle Hamiltonian $\hat{\mathcal{H}}(k,\phi)$ \cite{supply}, which is also periodic in $\phi$. Thus, we can define the energy band in the synthetic space $\mathbf{k}=(k,\phi)$ with the first Brillouin zone (BZ) $k\in(-\pi/l,\pi/l]$ and $\phi\in(-\pi/2,\pi/2]$. The topology of each band is characterized by the Chern number
\begin{equation}
\label{eq:eq4}
\mathcal{C}_n = \frac{1}{2\pi}\int_\mathrm{BZ}\mathcal{B}_n(\mathbf{k})d^2\mathbf{k},
\end{equation}
where $\mathcal{B}_n(\mathbf{k})=i\left(\langle\partial_\phi u_n|\partial_k u_n\rangle-\mathrm{c.c.}\right)$ is the Berry curvature of the $n$-th band, and $|u_n\rangle$ is the eigen state of $\hat{\mathcal{H}}(k,\phi)$. For the system considered above, we find three gapped bands with respective nontrivial topological numbers $\mathcal{C}_1,\mathcal{C}_2,\mathcal{C}_3=\{1,-2,1\}$, as shown in Fig.~\ref{fig:fig2}(b).

In this case, we can implement Thouless pumping \cite{thouless1983quantization}, while the parameter $\phi(t)$ is slowly modulated in time $t$. After a pumping cycle in which $\phi$ changes by $\pi$, the Hamiltonian returns to its initial form. If an energy band is filled or homogeneously populated, the mean displacement $\langle x\rangle$ of the exciton after one pumping cycle is quantized in units of lattice constant, i.e., $\langle x\rangle/l=\mathcal{C}_n$. For our system, the energy gap ${\Omega^2V(d)}/{8\Delta\left[\Delta+V(d)\right]}$ between the upper and the middle band is about two orders of magnitude larger than the gap between the middle and the lower band. Thus, to achieve better adiabaticity, we consider motion of the exciton within the upper band. As indicated by the pumping sequence shown in Fig.~\ref{fig:fig2}(c), we first shine a resonant field on sites $C_j$ and $A_{j+1}$ to produce an entangled state $|\psi_j\rangle=(1/\sqrt{2})(\hat{c}^\dagger_j+\hat{a}^\dagger_{j+1})|0\rangle$ using Rydberg blockade. With such an initialization and $\phi(0)=0$, we create an equally weighted Bloch states for the upper band. Then, we adiabatically ramp $\phi$ from $0$ to $\pi$, and observe the position of the exciton. As shown in Fig.~\ref{fig:fig2}(d), the mean displacement of the exciton after one pumping cycle is indeed $\langle x\rangle\approx 3d = l$, in agreement with the topology of the upper band. Since this energy band is almost flat in the $k$-dimension, such a quantized motion indicates a high-efficiency entangled state transfer from $|\psi_j\rangle$ to $|\psi_{j+1}\rangle$. It is worth pointing out that during the long pumping cycle $T$, the NNN interaction also comes into play. In fact, the long-range interaction induced NNN hoppings $\sum_{i}\left(J_A^\prime\hat{a}^\dagger_i\hat{c}_{i}+J_B^\prime\hat{b}^\dagger_i\hat{a}_{i+1}+J_C^\prime\hat{c}^\dagger_i\hat{b}_{i+1}+\mathrm{H.c.}\right)$ and the modifications to on-site potential can be viewed as perturbations to $\hat{H}_\mathrm{eff}$. We find that although these perturbations can significantly modify the spread $\langle x^2\rangle$ of the exciton, they do not change the mean displacement $\langle x\rangle$ \cite{supply}. Such a robust center-of-mass (COM) motion is protected by the topology of the band, which is invariant under continuous deformation of the Hamiltonian \cite{niu1984quantised,mei2018topology}.

{\it Example~2}.---The strong and nonlocal exciton-exciton interaction in the proposed system also makes it feasible for studying correlated transport \cite{fukuhara2013microscopic,preiss2015strongly}. Here, we consider the dynamics of two excitons in a homogeneously dressed ($\Delta_i=\Delta$ and $\Omega_i=\Omega$) chain with $V(2d)\ll|\Delta|$.

\begin{figure}
\centering
\includegraphics[width=\linewidth]{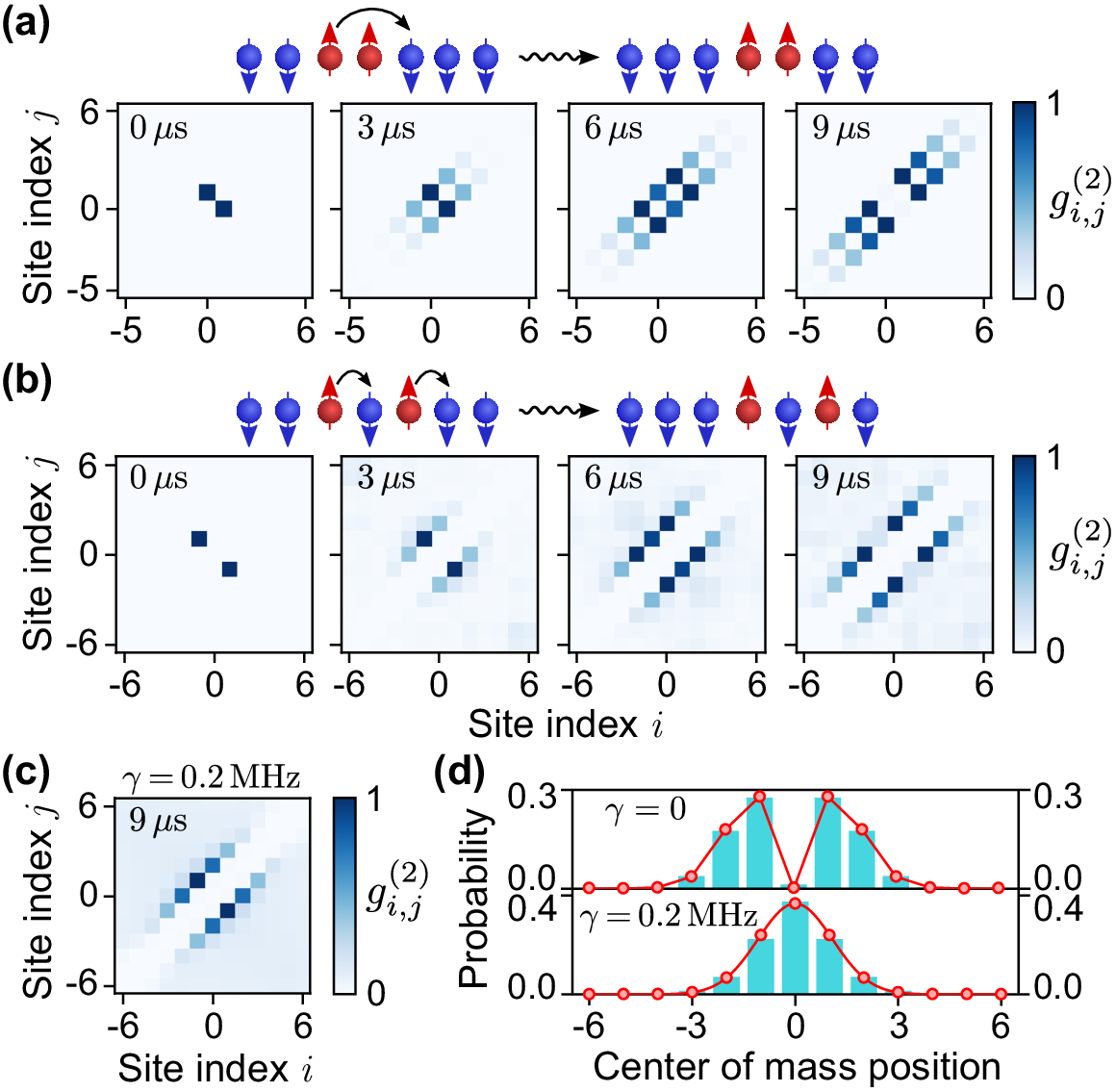}
\caption{(a) Evolution of the density-density correlation $g_{i,j}^{(2)}$ for the dimer state $|\Psi_{i,i+1}\rangle$. (b) Evolution of the density-density correlation for the high-order bound state $|\Psi_{i,i+2}\rangle$. (c) Correlation function at $9~\mu\mathrm{s}$, with the same initial state in (b) and $\gamma=0.2$~MHz. (d) Nomalized probability distribution of the COM position. The blue bars are numerical results obtained from the exact model with projective measurement, and the red dot-lines are fitted curves using Bessel function of the first kind (upper) and Gaussian function (lower), respectively. The parameters used are $\Omega/2\pi=5$~MHz, $\Delta/2\pi=-400$~MHz, $V(d)=-1.1\Delta$, $N=12$ in (a), and $\Delta/2\pi=30$~MHz, $V(d)=3\Delta$, $N=13$ in (b)-(d). The correlation functions are normalized to the maximal value.}
\label{fig:fig3}
\end{figure}

We first consider the dynamics of the dimer state $|\Psi_{i,i+1}\rangle$, which can be prepared via anti-blockade excitation satisfying $2\Delta+V(d)=0$ \cite{amthor2010}. As explained previously, such a tightly bound state can migrate through NNN hopping [see the upper panel of Fig.~\ref{fig:fig3}(a)], the hopping rate of which can be significant near the facilitation \cite{marcuzzi2017facilitation} region $\Delta+V(d)=0$. This correlated transport can be measured by the second-order correlation function $g^{(2)}_{i,j}=\langle \hat{a}^\dagger_i\hat{a}^\dagger_j\hat{a}_j\hat{a}_i\rangle$. As shown in Fig.~\ref{fig:fig3}(a), we find $g^{(2)}_{i,j}$ rapidly spreads on the diagonals $j=i\pm1$ while remains localized on the orthogonal directions, which confirms the existence of such a mobile bound state.

For excitons separated by more than one site, the state evolution is governed by $\hat{H}^\prime_\mathrm{eff}$. Unlike spin systems reported earlier \cite{fukuhara2013microscopic,letscher2018mobile}, the long-range interaction $U_{ij}$ can be tuned much larger than the exchange rate $J_{ij}$ here, which results in a highly anisotropic $XXZ$ model and permits the existence of high-order bound states. If the NNN interaction $U_{i,i+2}$ is sufficiently larger than the NN hopping rate $J_{i,i+1}$, excitons separated by one lattice site also forms bound state and exhibits correlated motion. Different from the dimer state $|\Psi_{i,i+1}\rangle$, transport of the high-order bound state $|\Psi_{i,i+2}\rangle$ relies on a second-order process with hopping rate $\sim J_{i,i+1}^2/U_{i,i+2}$ [see the upper panel of Fig.~\ref{fig:fig3}(b)]. As verified by numerical results shown in Fig.~\ref{fig:fig3}(b), $g^{(2)}_{i,j}$ also localizes on the skew diagonals, but spreads on the $j=i\pm2$ diagonals in this case. The influence of dephasing on this highly nonlocal correlated behavior is also investigated. In Fig.~\ref{fig:fig3}(c), we calculate the correlation function for $t>1/\gamma$, with $\gamma$ the dephasing rate introduced previously. Interestingly, the strong bunching of $g_{i,j}^{(2)}$ along $i=j\pm2$ survives, although its distribution is different from the coherent case. The small and uniform distribution of $g_{i,j}^{(2)}$ for $|i-j|>2$ indicates the diffusion equilibrium for unpaired free excitons has been established, while the strong NN interaction $U_{i,i+1}$ forbids the diffusion into $|i-j|=1$ region. To gain a deeper insight into such a correlated transport, we investigate the COM motion of two excitons. For coherent transport, this motion is characterized by a quantum random walk \cite{mulken2011continuous}, with a density distribution described by the Bessel function [see the upper panel of Fig.~\ref{fig:fig3}(d)]. In contrast, the COM density distribution at $t>1/\gamma$ is well fitted by a Gaussian function [see the lower panel of Fig.~\ref{fig:fig3}(d)]. This indicates that the high-order bound state we study exhibits diffusive expansion as a composite, which does not reach equilibrium due to its reduced diffusion rate compared to free excitons.

In conclusion, we propose a Rydberg-atom system for studying quantum transport dynamics, utilizing synthetic spin-exchange induced by vdW interaction. Our scheme does not require resonant DDI or off-diagonal vdW interaction, and thus avoids the complicated excitation schemes in multi-Rydberg-level systems. For the state-of-the-art experimental setup, its typical Rydberg lifetime $\sim~50~\mu\mathrm{s}$ \cite{levine2018high} will not hinder the exciton transport we discuss, and its influence on the dynamics can be eliminated by using projective measurement \cite{supply}. In addition to simulating quantum transport phenomena, this work opens up an avenue towards constructing exotic spin models with Rydberg atoms, which for instance can facilitate the study of many-body localization \cite{ponte2015many,smith2016many}.

\begin{acknowledgments}
This work is supported by the National Key R$\&$D Program of China (Grant No.~2018YFA0306504) and by NSFC (Grant No.~11804181). F.~Yang acknowledges valuable discussions with Prof.~K.~Ohmori, Prof.~S.~Sugawa, Dr.~J.~Yu and Dr.~F.~Reiter.
\end{acknowledgments}

\bibliography{main_text}

\end{document}


\preprint{APS/123-QED}

\title{Supplementary Material for ``Quantum Transport of Rydberg Exciton with Synthetic Spin-Exchange Interactions''}
\author{Fan Yang}
\affiliation{State Key Laboratory of Low Dimensional Quantum Physics, Department of Physics, Tsinghua University, Beijing 100084, China}
\author{Shuo Yang}
\affiliation{State Key Laboratory of Low Dimensional Quantum Physics, Department of Physics, Tsinghua University, Beijing 100084, China}
\author{Li You}
\affiliation{State Key Laboratory of Low Dimensional Quantum Physics, Department of Physics, Tsinghua University, Beijing 100084, China}
\affiliation{Beijing Academy of Quantum Information Sciences, Beijing 100193, China}
\maketitle
\onecolumngrid

This supplementary provides some technical details of the main text, including: (i) derivation of the many-body effective Hamiltonian (Sec.~\ref{sec:sec1}); (ii) discussion on exciton transport dynamics in a realistic system (Sec.~\ref{sec:sec2}); (iii) details of topological exciton pumping illustrated in the main text (Sec.~\ref{sec:sec3}).

\section{many-body quasi-degenerate perturbation analysis}\label{sec:sec1}
We first consider the effective Hamiltonian for a single Rydberg exciton. The initial Hamiltonian of the model system considered [see Eq.~(1) in the main text] can be decomposed as $\hat{H}=\hat{H}_0+\hat{V}$, where
\begin{equation}
\hat{H}_0 = \sum_{i}\Delta_i|\Psi_i\rangle\langle\Psi_i|+\sum_{i<j}(\Delta_i+\Delta_j+V_{ij})|\Psi_{ij}\rangle\langle \Psi_{ij}| + \hat{H}_\epsilon,
\quad \hat{V} = \sum_{i}\left(\frac{\Omega_i}{2}|g_i\rangle\langle r_i|+\frac{\Omega_i}{2}|r_i\rangle\langle g_i|\right),\label{eq:eq1}
\end{equation}
denote the diagonal free Hamiltonian and the off-diagonal laser-driving term, respectively, with $|\Psi_{i}\rangle=|g_1\cdots r_i\cdots g_N\rangle$ the single-exciton state, $|\Psi_{ij}\rangle=|g_1\cdots r_i\cdots r_j\cdots g_N\rangle$ the two-exciton state, $V_{ij}$ the van der Waals (vdW) interaction between atoms in state $|r_i\rangle$ and $|r_j\rangle$, and $\hat{H}_\epsilon$ the residue Hamiltonian for states containing more than two excitons.

In the large detuning regime ($\Delta_i\sim\Delta$, $|\Delta|\gg\Omega$), the singly excited state set $\{|\Psi_i\rangle,i=1,2,\cdots,N\}$ forms a quasi-degenerate subspace $\Pi_1$, the dynamics inside which can be described by applying second-order van Vleck perturbation theory \cite{shavitt1980quasidegenerate} to the exact Hamiltonian. To proceed the perturbation analysis, we first introduce the projection operator of the quasi-degenerate space ($\hat{\Pi}_g$) and the residue space ($\hat{\Pi}_r$), i.e.,
\begin{align}
\hat{\Pi}_{g} = \sum_{i}|\Psi_i\rangle\langle\Psi_i|,\quad
\hat{\Pi}_{r} = |G\rangle\langle G|+\sum_{i<j}|\Psi_{ij}\rangle\langle \Psi_{ij}|+\hat{\Pi}_{\epsilon},\label{eq:eq2}
\end{align}
where $|G\rangle=|g_1g_2\cdots g_N\rangle$ and $\hat{\Pi}_\epsilon$ denotes the projection operator for states containing more than two excitons. The coupling between the quasi-degenerate space and the residue space is governed by $\hat{V}^\prime = \hat{\Pi}_g\hat{V}\hat{\Pi}_r + \hat{\Pi}_r\hat{V}\hat{\Pi}_g$, which transforms single-exciton states into
\begin{equation}
\hat{V}^\prime|\Psi_j\rangle =\sum_{j<l}\frac{\Omega_l}{2}|\Psi_{jl}\rangle+\sum_{k<j}\frac{\Omega_k}{2}|\Psi_{kj}\rangle+\frac{\Omega_j}{2}|G\rangle,\quad
\langle\Psi_i|\hat{V}^\prime =\sum_{i<l}\frac{\Omega_l}{2}\langle \Psi_{il}|+\sum_{k<i}\frac{\Omega_k}{2}\langle \Psi_{ki}|+\frac{\Omega_i}{2}\langle G|.\label{eq:eq3}
\end{equation}
The elements of the second-order perturbative Hamiltonian $\hat{\mathcal{H}}^{(2)}$ inside space $\Pi_1$ is given by
\begin{align}
\langle\Psi_i|\hat{\mathcal{H}}^{(2)}|\Psi_j\rangle = \frac{1}{2}\langle\Psi_i|\hat{V}^\prime\frac{\hat{\Pi}_r}{\Delta_i-H_0}\hat{V}^\prime
+\hat{V}^\prime\frac{\hat{\Pi}_r}{\Delta_j-H_0}\hat{V}^\prime|\Psi_j\rangle.\label{eq:eq4}
\end{align}
With Eqs.~(\ref{eq:eq3}) and (\ref{eq:eq4}), the off-diagonal elements of $\hat{\mathcal{H}}^{(2)}$ are given by
\begin{equation}
\langle \Psi_i|\hat{\mathcal{H}}^{(2)}|\Psi_j\rangle=\frac{\Omega_{i}\Omega_jV_{ij}}{8\Delta_j(\Delta_j+V_{ij})}+\frac{\Omega_{i}\Omega_jV_{ij}}{8\Delta_i(\Delta_i+V_{ij})}, \quad (i\neq j)
\end{equation}
and the diagonal elements of $\hat{\mathcal{H}}^{(2)}$ are
\begin{equation}
\label{eq:eq7}
\langle \Psi_i|\hat{\mathcal{H}}^{(2)}|\Psi_i\rangle=\frac{\Omega_i^2}{2\Delta_i}-\sum_{j}\frac{\Omega_j^2}{4\Delta_j}
+\sum_{i<j}\frac{\Omega_j^2V_{ij}}{4\Delta_j(\Delta_j+V_{ij})}+\sum_{j<i}\frac{\Omega_j^2V_{ji}}{4\Delta_j(\Delta_j+V_{ji})}.
\end{equation}
The effective Hamiltonian $\hat{H}_\mathrm{eff}$ is then decomposed into $\hat{\mathcal{H}}^{(0)}+\hat{\mathcal{H}}^{(2)}$ with $\hat{\mathcal{H}}^{(0)}=\sum_{i}\Delta_i|\Psi_i\rangle\langle\Psi_i|$ the zeroth order term. Dropping the constant term $-\sum_{j}\Omega_j^2/4\Delta_j$ in Eq.~(\ref{eq:eq7}) and rewriting $\hat{H}_\mathrm{eff}$ in terms of the local spin operators, we can obtain the effective Hamiltonian [see Eq.~(2) in the main text],
\begin{equation}
\hat{H}_\mathrm{eff} =\sum_{i}\left(\Delta_i+\frac{\Omega_i^2}{2\Delta_i}\right)\hat{\sigma}_{rr}^i + \sum_{i\neq j} I_{ij}\hat{\sigma}_{rr}^i\hat{\sigma}_{gg}^j + J_{ij}\hat{\sigma}_{+}^i\hat{\sigma}_{-}^j,
\end{equation}
where $I_{ij}$ and $J_{ij}$ are Ising-type and spin-exchange interactions. Their explicit forms are given in the main text.

Next, we develop the effective theory involving mutiple Rydberg excitons. For the case of two excitons, we focus on the dynamics inside subspace $\Pi_2$ spanned by doubly excited state set $\{|\Psi_{ij}\rangle,i<j\}$. However, the bare energy $\Delta_i+\Delta_j + V_{ij}$ for state $|\Psi_{ij}\rangle$ is not quasi-degenerate for all $ij$, because the strong vdW interaction $V_{ij}$ can remove part of the degeneracy. As a result, $\Pi_2$ can be decomposed into several quasi-degenerate subspaces according to the value of $V_{ij}$, i.e., $\Pi_2=\Pi_{2}^{1}\bigcup\cdots\Pi_{2}^{m}\bigcup\cdots$, where the quasi-degenerate subspace $\Pi_{2}^{m}$ is spanned by $\{|\Psi_{ij}\rangle,|V_{ij}-E_m|\ll|\Delta|\}$ with $E_m$ the energy scale of $\Pi_{2}^{m}$ and $|E_{m}-E_{m^\prime}|\gtrsim|\Delta|$ for $m\neq m^\prime$. Up to second-order perturbations, these subspaces are decoupled with each other, and the perturbation Hamiltonian $\hat{\mathcal{H}}^{(2)}$ inside each subspace is given by ($k\neq i,j$)
\begin{align}
\langle \Psi_{ij}|\hat{\mathcal{H}}^{(2)}|\Psi_{ij}\rangle=\frac{\Omega_i^2}{2\Delta_i} + \frac{\Omega_j^2}{2\Delta_j} -\sum_{\mu}\frac{\Omega_\mu^2}{4\Delta_\mu}+
V_{ij}-(I_{ij}+I_{ji})+\sum_{\mu\neq i,j}\frac{\Omega_\mu^2(V_{i\mu}+V_{j\mu})}{4\Delta_\mu(\Delta_\mu+V_{i\mu}+V_{j\mu})},\label{eq:eq10}\\
\langle \Psi_{ik}|\hat{\mathcal{H}}^{(2)}|\Psi_{kj}\rangle=\langle \Psi_{ik}|\hat{\mathcal{H}}^{(2)}|\Psi_{jk}\rangle=
\langle \Psi_{ki}|\hat{\mathcal{H}}^{(2)}|\Psi_{kj}\rangle=\sum_{\beta=i,j}\frac{\Omega_i\Omega_jV_{ij}}{8(\Delta_\beta+V_{k\beta})(\Delta_\beta+V_{k\beta}+V_{ij})}.\label{eq:eq11}
\end{align}
For subspace $\Pi_2^{m}$ satisfying $E_m\ll|\Delta|$, the three-body interaction term in Eqs.~(\ref{eq:eq10}) and (\ref{eq:eq11}) can be approximated by a two-body interaction, i.e.,
\begin{equation}
\frac{\Omega_\mu^2(V_{i\mu}+V_{j\mu})}{4\Delta_\mu(\Delta_\mu+V_{i\mu}+V_{j\mu})}\approx\frac{\Omega_\mu^2V_{i\mu}}{4\Delta_\mu(\Delta_\mu+V_{i\mu})}
+\frac{\Omega_\mu^2V_{j\mu}}{4\Delta_\mu(\Delta_\mu+V_{j\mu})},\quad \frac{\Omega_i\Omega_jV_{ij}}{8(\Delta_\beta+V_{k\beta})(\Delta_\beta+V_{k\beta}+V_{ij})}\approx
\frac{\Omega_i\Omega_jV_{ij}}{8\Delta_\beta(\Delta_\beta+V_{ij})}.\nonumber
\end{equation}
With this approximation and $|\Delta|\gg |J_{ij}|$, the evolution of state $|\Psi_{ij}\rangle\in\Pi_2^{m}$ can be described by
\begin{equation}
\hat{H}_\mathrm{eff}^\prime =\sum_{i}\left(\Delta_i+\frac{\Omega_i^2}{2\Delta_i}\right)\hat{\sigma}_{rr}^i + \sum_{i\neq j} I_{ij}\hat{\sigma}_{rr}^i\hat{\sigma}_{gg}^j + J_{ij}\hat{\sigma}_{+}^i\hat{\sigma}_{-}^j+\sum_{i<j}(V_{ij}-I_{ij}-I_{ji})\hat{\sigma}^i_{rr}\hat{\sigma}^j_{rr},
\end{equation}
which is equivalent to the tight binding Hamiltonian $\hat{H}_\mathrm{eff}^\prime=\sum_i\mu_i\hat{a}^\dagger_i\hat{a}_i+\sum_{i<j}J_{ij}\left(\hat{a}^\dagger_i\hat{a}_{j}+ \hat{a}^\dagger_j\hat{a}_{i}\right)+U_{ij}\hat{a}^\dagger_i\hat{a}^\dagger_j\hat{a}_j\hat{a}_i$ given in the main text. This result can also be generalized to cases of multi-excitons, as long as the initial separation between excitons are large enough to ensure that the interactions between them are much smaller than the detuning, the effective dynamics are dominated by two-body interactions as described by $\hat{H}_\mathrm{eff}^\prime$.

However, if excitons are close to each other initially, they will form tightly bound state, the transport of which is dominated by many-body interactions. For example, in the two-exciton case, if the nearest-neighbor (NN) interaction $V_{i,i+1}$ is of the same order as $|\Delta|$, the dimer state $\{|\Psi_{i,i+1}\rangle,i=1,2,\cdots,N-1\}$ spans a quasi-degenerate subspace $\hat{\Pi}^\prime_2$. With Eqs.~(\ref{eq:eq10}) and (\ref{eq:eq11}), the dynamics inside $\hat{\Pi}^\prime_2$ are governed by $\hat{H}_\mathrm{eff}^\prime=\sum_i\epsilon_i|\Psi_{i,i+1}\rangle\langle\Psi_{i,i+1}|+ \left[J^{(2)}_i|\Psi_{i,i+1}\rangle\langle\Psi_{i+1,i+2}|+\mathrm{H.c.}\right]$, where $\epsilon_i=\Delta_i+\Delta_{i+1}+\langle\Psi_{i,i+1}|\hat{\mathcal{H}}^{(2)}|\Psi_{i,i+1}\rangle$ is the on-site potential, and
\begin{equation}
J^{(2)}_i=\frac{\Omega_i\Omega_{i+2}V_{i,i+2}}{8(\Delta_i+V_{i,i+1})(\Delta_i+V_{i+1,i}+V_{i,i+2})} +\frac{\Omega_i\Omega_{i+2}V_{i,i+2}}{8(\Delta_{i+2}+V_{i+1,i+2})(\Delta_{i+2}+V_{i+1,i+2}+V_{i,i+2})}
\end{equation}
is the spin-exchange interaction between the next-nearest-neighbor (NNN). In fact, $J^{(2)}_i$ is a three-body interaction associated with the process $J^{(2)}_i\hat{\sigma}^i_+\hat{\sigma}^{i+1}_{rr}\hat{\sigma}^{i+2}_-+\mathrm{H.c.}$, which differs from the two-body NNN exchange rate $J_{i,i+2}$.

\section{Transport dynamics in a realistic system}\label{sec:sec2}
In this section, we consider the effective exciton dynamics in a realistic system, where the dephasing, the spontaneous decay of the Rydberg state, and the atomic positional disorder could influence the exciton transport model established in the previous section.
\subsection{Effects of dephasing}

\begin{figure}
\centering
\includegraphics[width=0.8\linewidth]{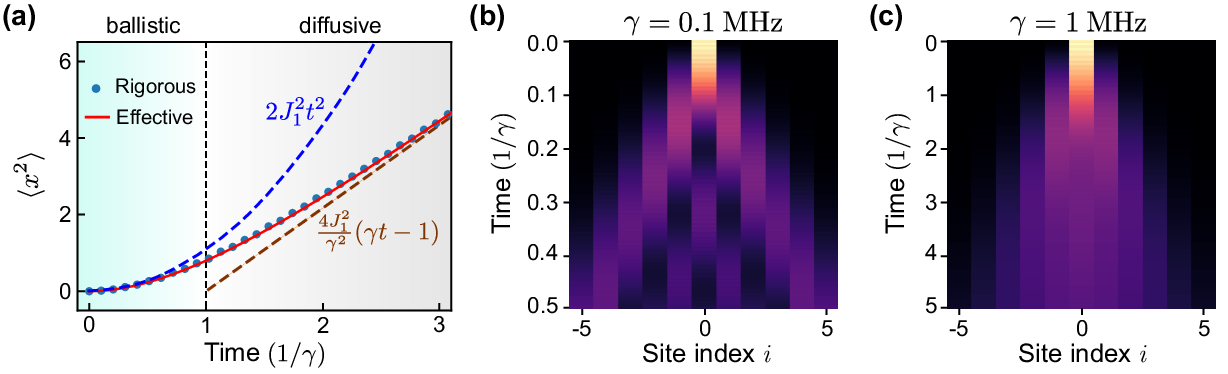}
\caption{(a) Evolution of the mean squared displacement. The blue dots are calculated by the exact model followed by projective measurements, and the red line represent the analytic solution. (b) and (c) show the evolution of exciton density distributions in the ballistic and diffusive regime, respectively, where a brighter color indicates a larger probability. The parameters used are $\Omega/2\pi=5$~MHz, $\Delta/2\pi=50$~MHz, $V(d)=3\Delta$ ($d=4.4~\mu\mathrm{m}$), $N=11$, and $\gamma=$ 0.8, 0.1, 1~MHz for (a), (b), (c).}\label{fig:fig4}
\end{figure}
For the state-of-the-art experimental system \cite{levine2018high}, the main decoherence channel is the pure dephasing caused by atomic Doppler effect and phase noises of the dressing laser. Thus, we first consider the open-system dynamics with local dephasings. In this case, the evolution of system density matrix $\hat{\rho}$ can be described by the master equation $\partial_t{\hat{\rho}} =-i[\hat{H},\hat{\rho}]+\sum_{i}\mathcal{L}[\sqrt{\gamma}\hat{\sigma}_{rr}^{i}]\hat{\rho}$ with the Lindblad operator $\mathcal{L}[\hat{\sigma}]\hat{\rho}=\hat{\sigma}\hat{\rho}\hat{\sigma}^\dagger- \frac{1}{2}(\hat{\sigma}^\dagger\hat{\sigma}\hat{\rho}+\hat{\rho}\hat{\sigma}^\dagger\hat{\sigma})$ \cite{lesanovsky2013kinetic}. Before discussing exciton transport, we discuss the influence of dephasing on a single atom. In the weak-dephasing regime ($|\Delta|\gg\gamma$), the equation of motion for $\rho_{rr} = \langle r|\hat{\rho}|r\rangle$ is governed by
\begin{equation}
\frac{d}{dt}\rho_{rr} = -\Gamma\left(\rho_{rr}-\frac{1}{2}\right),\quad\mathrm{with}\quad \Gamma=\frac{\Omega^2\gamma/2}{\Delta^2+(\gamma/2)^2}\approx
\frac{\Omega^2\gamma}{2\Delta^2},\label{eq:eq12}
\end{equation}
which indicates that for an excitation-free initial state $|g\rangle$ [$\rho_{rr}(0)=0$], the Rydberg exciton grows with a rate $\Gamma/2$. Similarly, the excited initial state $|r\rangle$ has a loss rate $\Gamma/2$. In multi-atom systems, such an exciton growth (loss) can be suppressed by vdW interactions \cite{lesanovsky2013kinetic}, but the growth (loss) rate can still be roughly estimated as $\Gamma/2$.

For simulating exciton transport dynamics, the total exciton number should be conserved during the state evolution. Thus, we focus on the dynamics in times shorter than $t_c=\min\left\{\Delta_i^2/\gamma\Omega_i^2\right\}$, during which the exciton growth (loss) event rarely occurs and can be subtracted out by projective measurement. In this regime, the dynamics can be effectively described by $\partial_t{\hat{\rho}} =-i[\hat{H}_\mathrm{eff}^\prime,\hat{\rho}]+\sum_{i}\mathcal{L}[\sqrt{\gamma}\hat{a}_i^\dagger\hat{a}_i]\hat{\rho}$, which conserves the total exciton number $\sum_i\hat{a}^\dagger_i\hat{a}_i$.

To see the crossover between coherent and incoherent exciton motion, we consider a simple case: transport of a single excitation in a homogenously dressed infinite chain described by $\hat{H}_\mathrm{eff} = \sum_{i,d}J_d\left({\hat{a}_i^\dagger\hat{a}_{i+d}+\hat{a}_{i+d}^\dagger\hat{a}_{i}}\right)$, where the hopping rate $J_d=J_{i,i+d}$ only depends on the distance $d>0$. The density matrix for a single exciton can be expressed as $\hat{\rho}=\sum_{m,n}\rho_{m,n}\hat{a}^\dagger_m|0\rangle\langle 0|\hat{a}_n$, and the resulting equation of motion is equivalent to the Haken-Reineker-Strobl model with cohorent hoppings and on-site dephasings:
\begin{equation}
\label{eq:eq13}
\dot{\rho}_{m,n} = -i\sum_{d>0}J_d\left(\rho_{m+d,n}+\rho_{m-d,n}-\rho_{m,n-d}-\rho_{m,n+d}\right)-\gamma(1-\delta_{mn})\rho_{m,n}.
\end{equation}
For the initial state $\hat{\rho}=\hat{a}^\dagger_0|0\rangle\langle 0|\hat{a}_0$, the equation of motion for the mean square displacement $\langle x^2\rangle = \sum_{n}n^2\rho_{n,n}$ is
\begin{equation}
\frac{d^2}{dt^2} \langle x^2\rangle + \gamma\frac{d}{dt} \langle x^2\rangle=4\sum_d(dJ_d)^2, \quad \mathrm{with} \quad \left.\frac{d\langle x^2\rangle}{dt}\right|_{t=0}=0,\
 \left.\langle x^2\rangle\right|_{t=0}=0.\label{eq:eq14}
\end{equation}
The solution of Eq.~(\ref{eq:eq14}) is given by
\begin{equation}
\langle x(t)^2\rangle =\frac{4\sum_d(dJ_d)^2}{\gamma^2}\left(\gamma t+e^{-\gamma t}-1\right)\approx
\left\{\
\begin{matrix}
\left[2\sum_d(dJ_d)^2\right]t^2, \quad\mathrm{for}\quad t\ll\gamma^{-1},\\ \\ \left[4\sum_d(dJ_d)^2/\gamma\right] t, \quad\mathrm{for}\quad t\gg\gamma^{-1},
\end{matrix}
\right.\label{eq:eq15}
\end{equation}
which indicates that the transport experiences a crossover from ballistic spreading ($\sim t^2$) to diffusive expansion ($\sim t$) as the system evolves, with the characteristic time being $1/\gamma$. Such a crossover is verified by numerical simulations shown in Fig.~\ref{fig:fig4}(a). The transport process is coherent in the ballistic spreading regime [Fig.~\ref{fig:fig4}(b)], as characterized by the interference fringes during the spreading. In the diffusive expansion regime, the exciton motion exhibits incoherent features, as indicated by the Gaussian type density distributions [Fig.~\ref{fig:fig4}(c)].

\subsection{Effects of spontaneous radiative decay}
While the dephasing can significantly modify the transport dynamics, the influence of finite Rydberg lifetime is negligible. This is because for the current experimental system \cite{levine2018high}, the Rydberg state lifetime $\tau_r\sim 50~\mu\mathrm{s}$ is sufficiently long for observing exciton transport at a rate $\sim 1~\mathrm{MHz}$. Further, the influence of spontaneous decay can be removed from final results by using the projective measurement introduced in the main text, as we show below.

The full open-system dynamics including both dephasing and spontaneous decay can be described by the master equation $\partial_t{\hat{\rho}} =-i[\hat{H},\hat{\rho}]+\sum_{i}\mathcal{L}[\sqrt{\gamma}\hat{\sigma}_{rr}^{i}]\hat{\rho}+\sum_{i}\mathcal{L}[\sqrt{\kappa}\hat{\sigma}_{-}^{i}]\hat{\rho}$, where $\kappa$ is the spontaneous decay rate. In this case, the density matrix can be written as $\hat{\rho}=\sum_{m,n}\rho_{m,n}\hat{a}^\dagger_m|0\rangle\langle 0|\hat{a}_n+\hat{\rho}^{\prime}$, where $\hat{\rho}^\prime$ represents the density matrix when spontaneous decay occurs. The evolution of the matrix element $\rho_{m,n}$ is then given by
\begin{equation}
\label{eq:eq16}
\dot{\rho}_{m,n} = -\kappa\rho_{m,n}-i\sum_{d>0}J_d\left(\rho_{m+d,n}+\rho_{m-d,n}-\rho_{m,n-d}-\rho_{m,n+d}\right)-\gamma(1-\delta_{mn})\rho_{m,n}.
\end{equation}
Comparing Eq.~(\ref{eq:eq16}) with Eq.~(\ref{eq:eq13}), we find that the spontaneous decay simply contributes an exponential decay factor $e^{-\kappa t}$ to the elements $\rho_{m,n}$. With the projective measurement, the events described by $\hat{\rho}^\prime$ are discarded, and the refined density matrix is just the same as that described by the dephasing model. This is verified by numerical results shown in Fig.~\ref{fig:fig5}, where we calculate the mean squared displacement $\langle x^2\rangle$ for the exciton initially localized at the center of a periodic chain containing 7 sites. As shown in Fig.~\ref{fig:fig5}(a), the only difference between the dephasing model and the full model is simply an exponential factor. By using the projective measurement, the full model has the same prediction as the dephasing model, as proved by Fig.~\ref{fig:fig5}(b).

\subsection{Influences of positional disorders}
\begin{figure}
\centering
\includegraphics[width=0.75\linewidth]{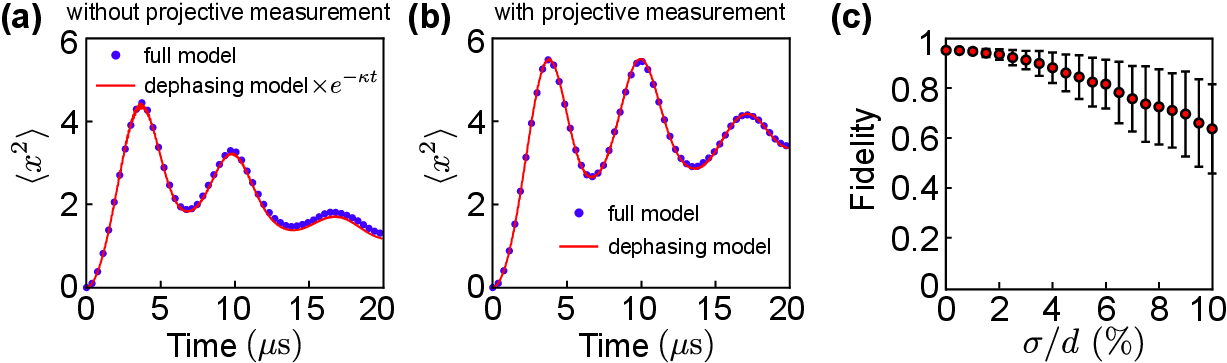}
\caption{(a) and (b) show the evolution of $\langle x^2\rangle$, where (b) and (a) are calculated with and without projective measurement, respectively. The parameters are $\gamma=0.1~$MHz, $\kappa=0.05~$MHz, $\Omega/2\pi=5$~MHz, $\Delta=10\Omega$ and $V(d)=3\Delta$. (c) Fidelity of the entanglement transfer for different uncertainties $\sigma$, with other parameters the same as Fig.~1(d) in the main text. The data points are averaged over 500 calculations, and the error bars mark one standard deviation intervals.}\label{fig:fig5}
\end{figure}
For the current experiment, the finite temprature of the atom induces an uncertainty of its position, resulting in an uncertainty of the vdW interaction $V_{ij}$.
In the facilitation regime discussed in Ref.~\cite{marcuzzi2017facilitation}, the exciton transport requires the fluctuation of the NN interaction much smaller than the Rabi frequency, i.e., $\delta V_{i,i+1}\ll\Omega$, otherwise the facilitation condition $\Delta+V_{i,i+1}=0$ is broken and the initial excitation becomes highly localized. In our dressing scheme, both the on-site potential and the exchange rate are determined by the effective interaction strength $J_{ij}$, such that the influence of the disorder is determined by its fluctuation $\delta J_{ij}$. If the ratio $\delta J_{ij}/J_{ij}=\frac{\Delta(\delta V_{ij})}{(\Delta+V_{ij})V_{ij}}\sim\delta V_{ij}/V_{ij}\leq\delta V_{i,i+1}/V_{i,i+1}$ is much smaller than unity, the positional disorder will not hinder the transport process. Due to the fact $V_{i,i+1}\gg\Omega$, the condition $\delta V_{i,i+1}\ll V_{i,i+1}$ is much looser than $\delta V_{i,i+1}\ll\Omega$, which suggests that the transport in our scheme is more robust against positional disorders compared with the facilitation regime.

The robustness of our scheme is verified by the entanglement transfer protocol shown in Fig.~1(d) of the main text. In the calculation, we consider the Gaussian distribution of the atom position along the chain direction with an uncertainty $\sigma$. For vdW interaction $V_{ij}=C_6/r_{ij}$, we have $\delta V_{ij}/V_{ij}=6\delta r_{ij}/r_{ij}\leq6\sqrt{2}\sigma/d$. The fidelities of the entanglement transfer for different values of $\sigma/d$ are shown in Fig.~\ref{fig:fig5}(c). For a realistic $\sigma=0.1~\mu\mathrm{m}$ \cite{marcuzzi2017facilitation}, the ratio $\sigma/d\approx2\%$ ($\delta V_{ij}/V_{ij}<0.2$) is small enough to ensure a high transport efficiency.

\section{topological exciton transport}\label{sec:sec3}
With the NN interacting approximation, the system for implementing topological exciton pumping can be mapped to the generalized Rice-Mele model [see Eq.~(3) in the main text] with
\begin{equation}
\hat{H}_\mathrm{eff}(\phi)=\sum_{i}\left[J_A(\phi)\hat{a}^\dagger_i\hat{b}_{i}+J_B(\phi)\hat{b}^\dagger_i\hat{c}_{i}+J_C(\phi)\hat{c}^\dagger_i\hat{a}_{i+1}+\mathrm{H.c.}\right]
+\sum_i\left[\mu_A(\phi)\hat{a}^\dagger_i\hat{a}_i+\mu_B(\phi)\hat{b}^\dagger_i\hat{b}_i+\mu_C(\phi)\hat{c}^\dagger_i\hat{c}_i\right].
\end{equation}
The hopping rates and the on-site potentials (after dropping the constant term $\Delta$) are given by
\begin{align}
J_A(\phi) = U\sin^2(\phi+\pi/4)\sin^2(\phi),\quad
\mu_A(\phi) = E\sin^4(\phi+\pi/4)+U\left[\sin^4(\phi)+\sin^4(\phi-\pi/4)\right],\\
J_B(\phi) = U\sin^2(\phi)\sin^2(\phi-\pi/4),\quad
\mu_B(\phi) = E\sin^4(\phi)+U\left[\sin^4(\phi-\pi/4)+\sin^4(\phi+\pi/4)\right],\\
J_C(\phi) = U\sin^2(\phi-\pi/4)\sin^2(\phi+\pi/4),\quad
\mu_C(\phi) = E\sin^4(\phi-\pi/4)+U\left[\sin^4(\phi+\pi/4)+\sin^4(\phi)\right],
\end{align}
with $E=\Omega^2/2\Delta$ and $U=\Omega^2V(d)/4\Delta[\Delta+V(d)]$. Transforming real-space exciton operators into quasi-momentum $k$-space with $k\in(-\pi/l,\pi/l]$ and
\begin{equation}
\hat{a}^\dagger_k = \frac{1}{\sqrt{N}}\sum_{j=1}^N\hat{a}^\dagger_je^{ikjl}, \quad \hat{b}^\dagger_k = \frac{1}{\sqrt{N}}\sum_{j=1}^N\hat{b}^\dagger_je^{ikjl}, \quad
\hat{c}^\dagger_k = \frac{1}{\sqrt{N}}\sum_{j=1}^N\hat{c}^\dagger_je^{ikjl},
\end{equation}
the effective Hamiltonian can be decomposed into $\hat{H}_\mathrm{eff}(\phi)=\sum_k\hat{\mathcal{H}}(k,\phi)$ with
\begin{equation}
\hat{\mathcal{H}}(k,\phi) =
\begin{pmatrix}
\hat{a}^\dagger_k&\hat{b}^\dagger_k&\hat{c}^\dagger_k
\end{pmatrix}
\begin{pmatrix}
\mu_A(\phi)&J_A(\phi)&J_C(\phi)e^{-ikl}\\
J_A(\phi)&\mu_B(\phi)&J_B(\phi)\\
J_C(\phi)e^{ikl}&J_B(\phi)&\mu_C(\phi)
\end{pmatrix}
\begin{pmatrix}
\hat{a}_k \\ \hat{b}_k \\ \hat{c}_k
\end{pmatrix}.
\end{equation}
The single-particle eigen state of $\hat{\mathcal{H}}(k,\phi)$ satisfying $\hat{\mathcal{H}}(k,\phi)|u_n(k,\phi)\rangle=E_n(k,\phi)|u_n(k,\phi)\rangle$ can be expressed as
\begin{equation}
|u_n(k,\phi)\rangle =
\begin{pmatrix}
\hat{a}^\dagger_k|0\rangle&\hat{b}^\dagger_k|0\rangle&\hat{c}^\dagger_k|0\rangle
\end{pmatrix}
\begin{pmatrix}
u_{n,a}(k,\phi) \\ u_{n,b}(k,\phi) \\ u_{n,c}(k,\phi)
\end{pmatrix},
\end{equation}
where $n$ denotes the band index. Since $\hat{\mathcal{H}}(k,\phi)$ is periodic in both $k$ and $\phi$, we can introduce the energy band in the two-dimensional space spanned by $\mathbf{k}=(k,\phi)$ and calculate its topologic number according to Eq.~(4) of the main text. For an initial state $|\psi(0)\rangle=(1/\sqrt{N})\sum_ke^{i\xi(k)}|u_n(k,\phi_0)\rangle$ that homogeneously populates [$\xi(k)\in \mathbb{R}$] each $k$ component of the $n$-th band, the state evolution in the adiabatic limit is given by $|\psi(t)\rangle=(1/\sqrt{N})\sum_k|\psi_k(t)\rangle$, with
\begin{equation}
|\psi_k(t)\rangle = \exp\left\{i\xi(k)-i\int_0^tdt^\prime E_n[k,\phi(t)]-\int_{\phi_0}^{\phi(t)}d\phi\langle u_n(k,\phi)|\partial_\phi|u_n(k,\phi)\rangle\right\}|u_n(k,\phi)\rangle.
\end{equation}
The mean position of the exciton in the continuous limit is expressed as
\begin{equation}
\langle X(t)\rangle = \frac{l}{2\pi}\int_{-\pi/l}^{\pi/l}dk\langle\psi_k(t)|i\partial_k|\psi_k(t)\rangle +dP_b(t) +2dP_c(t),
\end{equation}
where $P_b$ and $P_c$ denote the exciton distribution probability at site $B$ and $C$, respectively. After one pumping cycle $T$ with $\phi(T)=\phi_0+\pi$, the mean displacement $\langle x(t)\rangle=\langle X(T)\rangle-\langle X(0)\rangle$ is determined by
\begin{equation}
\langle x(T)\rangle = \frac{l}{2\pi}\int_{\phi_0}^{\phi_0+\pi}d\phi\int_{-\pi/l}^{\pi/l}dk \ i\left[\langle\partial_{\phi}u_n(k,\phi)|\partial_ku_n(k,\phi)\rangle-\mathrm{c.c.}\right]=l\mathcal{C}_n,
\end{equation}
which proves that the center of mass (COM) motion of the exciton is quantized in units of the lattice constant.

Exact diagonalization of $\hat{\mathcal{H}}(k,\phi)$ reveals that the energy gap between the upper ($n=3$) and the middle band ($n=2$) is $U/2$, while the gap between the middle and the lower band ($n=1$) is $\sim2U\sin^4(\pi/8)-U^2/[32(E-U)\sin^4(3\pi/8)]\approx0.017U$ for the parameter adopted in the main text [$V(d)=3\Delta$]. To achieve a better adiabaticity, we consider the transport of the exciton within the upper band. To this end, we consider to prepare an initial state $|\psi_j\rangle=(1/\sqrt{2})(\hat{c}^\dagger_j+\hat{a}^\dagger_{j+1})|0\rangle$ by using Rydberg blockade, and initialize the pumping with $\phi(0)=0$ where the eigen states of $\hat{\mathcal{H}}(k,\phi)$ are $|u_1(k,0)\rangle=\hat{b}^\dagger_k|0\rangle$, $|u_2(k,0)\rangle=(1/\sqrt{2})(e^{-ikl}\hat{a}^\dagger_k-\hat{c}^\dagger_k)|0\rangle$, and $|u_3(k,0)\rangle=(1/\sqrt{2})(e^{-ikl}\hat{a}^\dagger_k+\hat{c}^\dagger_k)|0\rangle$, such that $|\psi_j\rangle=(1/\sqrt{N})\sum_ke^{-ikjl}|u_3(k,0)\rangle$ creates a homogeneous population of the upper band. Adiabatically ramping $\phi(t)$ from $0$ to $\pi$, we can observe the topological exciton transport within the upper band.

\begin{figure}
\centering
\includegraphics[width=\linewidth]{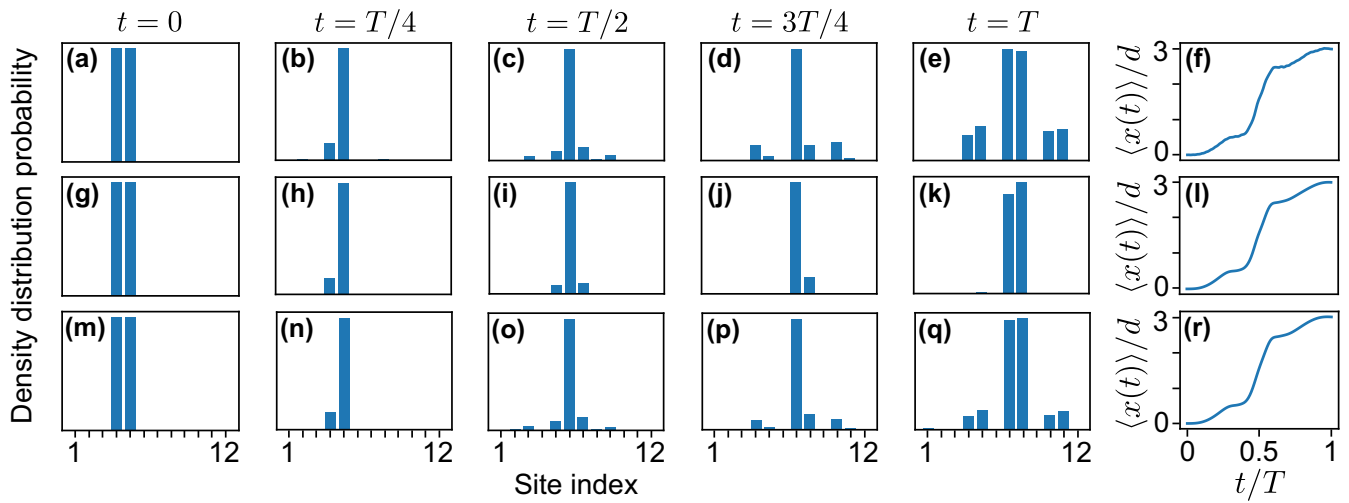}
\caption{(a) Evolution of the exciton density distribution probability and the mean displacement $\langle x(t)\rangle$. The parameters used are the same as in Fig.~2 of the main text.}\label{fig:fig6}
\end{figure}

In a more rigorous treatment, we need to take the vdW interaction between the NNN into consideration, which contributes a perturbation Hamiltonian
\begin{equation}
\Delta\hat{H}_\mathrm{eff}=\sum_{i}\left[J_A^\prime\hat{a}^\dagger_i\hat{c}_{i}+J_B^\prime\hat{b}^\dagger_i\hat{a}_{i+1}+J_C^\prime\hat{c}^\dagger_i\hat{b}_{i+1}+\mathrm{H.c.}\right]
+\sum_i\left[\Delta\mu_A\hat{a}^\dagger_i\hat{a}_i+\Delta\mu_B\hat{b}^\dagger_i\hat{b}_i+\Delta\mu_C\hat{c}^\dagger_i\hat{c}_i\right].
\end{equation}
The NNN hopping strength $J_{A}^\prime$, $J_{B}^\prime$, $J_{C}^\prime$, and the on-site potential modifications $\Delta\mu_{A}$, $\Delta\mu_{B}$, $\Delta\mu_{C}$ are of the order as $U^\prime=\Omega^2V(2d)/4\Delta[\Delta+V(2d)]\approx0.06U$. Although $\Delta \hat{H}_\mathrm{eff}$ is much smaller than $\hat{H}_\mathrm{eff}$, it has significant impact on the transport dynamics during the long pumping period $T\approx41U^{-1}$. As shown in Fig.~\ref{fig:fig6}, the evolution of exciton density distributions obtained with the exact model [Figs.~\ref{fig:fig6}(a)-\ref{fig:fig6}(e)] can be described by the NN approximate model [Figs.~\ref{fig:fig6}(g)-\ref{fig:fig6}(k)] only at early times. At large times, the spreading of excitations are much larger than that predicted by the NN model, but can be well described by the effective model including the NNN perturbation term $\Delta\hat{H}_\mathrm{eff}$ [Figs.~\ref{fig:fig6}(m)-\ref{fig:fig6}(q)]. Such a spreading indicates that the NNN perturbation can significantly modify the mean square displacement $\langle x^2\rangle$ of the exciton. Nevertheless, the mean displacement $\langle x\rangle$ is not influenced by this perturbation, which is almost the same for the exact model [Fig.~\ref{fig:fig6}(f)], the NN effective model [Fig.~\ref{fig:fig6}(l)], and the NNN effective model [Fig.~\ref{fig:fig6}(r)]. Such a robust COM motion is protected by the topology of the corresponding energy band.

\bibliography{supply}